# Intelligent Virtual Assistant knows Your Life

Hyunji Chung and Sangjin Lee

**Abstract**—In the IoT world, intelligent virtual assistant (IVA) is a popular service to interact with users based on voice command. For optimal performance and efficient data management, famous IVAs like Amazon Alexa and Google Assistant usually operate based on the cloud computing architecture. In this process, a large amount of behavioral traces that include user's voice activity history with detailed descriptions can be stored in the remote servers within an IVA ecosystem. If those data (as also known as IVA cloud-native data) are leaked by attacks, malicious person may be able to not only harvest detailed usage history of IVA services, but also reveals additional user-related information through various data analysis techniques. In this paper, we firstly show and categorize types of IVA-related data that can be collected from popular IVA, Amazon Alexa. We then analyze an experimental dataset covering three months with Alexa service, and characterize the properties of user's lifestyle and life patterns. Our results show that it is possible to uncover new insights on personal information such as user interests, IVA usage patterns and sleeping/wake-up patterns. The results presented in this paper provide important implications for and privacy threats to IVA vendors and users as well.

## 1 INTRODUCTION

THE Internet of Things (IoT) is evolving rapidly. Analysts predict that the worldwide IoT market will grow to $1.7 trillion in 2020 with a compound annual growth rate (CAGR) of 16.9% [1]. Gartner predicts that 25% of households using an intelligent virtual assistant (IVA) will have two or more devices by 2020 [2]. In the IoT world, an IVA is a popular service to communicate with users based on voice command. For example, an IVA can be embedded on smart speaker, smart refrigerators, connected car, etc.

For optimal performance and efficient data management, famous IVAs like Amazon Alexa and Google Assistant are usually based on the cloud computing. In this process, a large amount of behavioral traces that include user's voice activity history with detailed descriptions can be stored in the remote cloud servers within an IVA ecosystem. If those data are stolen or leaked by cyber attack such as data breach, malicious person may be able to not only harvest detailed usage history of IVA services, but also reveal additional user-related information through various data analysis techniques.

• *H. Chung and S. Lee are with Korea University, Seoul, South Korea. E-mails: foryou7187@korea.ac.kr, sangjin@korea.ac.kr*

In this paper, we handle an actual user's data stored in the most famous IVA, Amazon Alexa cloud. The highlights of this work are summarized as follows:

1) At first, this paper shows and categorize types of user generated Alexa data.
2) We analyze the experimental dataset covering several months with various Alexa-enabled devices such as Echo Dot, Dash Wand, Fire HD and Fire TV.
3) Through a variety of data analysis techniques, this paper shows that it is possible to uncover new insights on personal information such as interests and life patterns. Furthermore, the results of this work provide important implications for privacy risks to IVA vendors as well as end users.

The rest of the paper is organized as follows. We explain background knowledge in Section 2. We describe our data collection methodology and Alexa cloud-native datasets in Section 3. Through data analysis using visualization techniques, we characterize the properties of user's lifestyle and life patterns in Section 4. Based on our research results, Section 5 discusses implications for IVA users and vendors. Finally, we conclude in Section 6.

## 2 BACKGROUNDS

In the cloud computing environment, cloud is an environment of the hardware and software resources in the data centers that provide diverse services to satisfy the user's requirements. In the IoT world, for optimal performance and efficient data management, famous IoT products and services usually operate based on the cloud computing architecture. In this case, a large amount of user-related information can be stored in the remote cloud servers. In order to enhance security against cyber attacks, each segment of data is encrypted and separately distributed in the databases over the cloud [3]. However, data encryption is not enough for information security and data privacy. This is because most breaches happen when a hacker breaks into the application layer, exploits weak APIs, gets access to user credentials and steals sensitive data [4].

There have been serious data breaches reported in 2005 or beyond [5]. Firstly, more than 68 million Dropbox users had their usernames and passwords compromised at the end of August 2016. Secondly, 700 current and former Snapchat employees had their personal information, including names, social security numbers, and wage (or payroll) data. Thirdly, over 43 million Weebly users were stolen data, including usernames, passwords, e-mail addresses, and IP information in February. Fourthly, about 500 million Yahoo users had stolen information such as e-mail addresses, passwords, full user names, dates of birth, telephone numbers, and in some cases, security questions and answers in late 2014 [5].

According to Chung *et al.*, a large amount of behavioral traces that include user's activity history with detailed descriptions can be stored in the Alexa servers [6]. If those cloud-native data are leaked by cyber attacks, a hacker may be able to harvest the detailed usage history of Alexa services such as playing music, setting an alarm, checking traffic, asking a question, calling and messaging. Through simple data analysis techniques, hackers can reveal additional user-related information such as lifestyle and life pattern. Criminals can utilize stolen information to reveal when user usually wake up and go to bed. It means the hacker can monitor other's life on the remote side. Just in case user home address, information has also been leaked, a worst-case scenario where a cyber attacker turns to be a theft in the real world occurs.

In these days, studies on data privacy are essential parts for enhancing trust of various information technologies. Unfortunately, there has been little research reported on IVA cloud-native data from perspectives of data and user privacy. In this regard, Chung *et al.* pointed out security vulnerabilities and privacy risks of cloud-based IVA ecosystems [7]. As a follow-up research, this paper reports on analysis results with an experimental dataset from Amazon Alexa cloud, and characterize the properties of a user's lifestyle and life patterns.

## 3 MEASUREMENT METHODOLOGY

To obtain insights into user behaviors over IVA, we collected a variety of data from Amazon Alexa cloud. In this section, we describe about how we collected Amazon Alexa cloud data from its cloud, and then present several statistics on the data.

### 3.1 Cloud-based Amazon Alexa Ecosystem

As mentioned above, we focused on Amazon Alexa and its ecosystem for this work. There are various Alexa-enabled devices (Echo family, Dash Wand, Fire Tablet, Fire TV, etc.) that are required for interacting with the Alexa cloud service.

With Alexa IVA, users can perform various things, including but not limited to managing to-do lists, playing music, setting alarms, placing orders, searching information and checking traffic [8]. In this process, the Alexa cloud produces and stores various types of digital traces (logs) related to a user's behaviors.

### 3.2 Data Collection Methodology

Since we were interested in studying IVA-related user behavior analysis, we needed to collect user's activity history from IVA cloud as many as possible. This subsection briefly explains our methodology for collecting Alexa's cloud-native data.

Like other cloud services, the Alexa cloud also operates with pre-defined APIs to transceive data. In our previous research, we identified internal APIs that can be used to acquire cloud data from Alexa cloud [6]. Using these APIs that we identified, we automatically collected usage history logs stored in cloud side for supporting this study.

### 3.3 Data Description

The data we collected contains a participant's daily life during about three months together with a variety Alexa-related devices, including two Echo Dots, one

Dash Wand, one Fire Tablet and one Fire TV. As a result, we had Alexa's cloud-native data for the period of 121 consecutive days from April 1 to July 31, 2017. The collected dataset consisits of various types of data as listed in **Table 1**. Note that the data types are extracted by parsing actual log entries returned from Alexa cloud.

Table 1. Types of Alexa cloud-native data

| No. | Detected data type | Description |
|-----|--------------------|-------------|
| 1 | History | User activity history |
| 2 | Shopping itme | Shopping list |
| 3 | Task | To-do list |
| 4 | Eon card | Schedules |
| 5 | Notification card | Alarms and remainders |
| 6 | Puffin card | Restaurant information |
| 7 | Traffic card | Traffic information |
| 8 | Sport card | Sport information |

The data of 'History' type indicates a user's activity history. That is, each event occurred by a user produces a history entry for logging results of the event. Therfore, all other data types (No. 2 to 8 in **Table 1**) can be identified from the History data. Interestingly, the Alexa cloud also manages additional event records for each category of actions, such as 'Shopping item', 'Task', 'Eon card', 'Notification card', 'Puffin Card', 'Traffic card' and 'Sport card'. For reference, all the log entries listed in **Table 1** contain a UNIX timestamp, so it can be utilized for understanding a user's daily life more accurately.

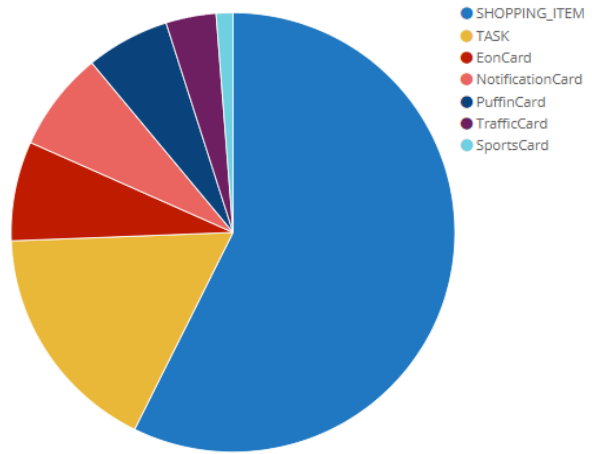

**Figure 1. Pie chart of data types detected from a dataset**

**Figure 1** tells us the ratio of detected data types. Totally, we gathered about 2,000 log entries from Alexa cloud. The figure shows distribution of the collected logs as follows: shopping item (57.32%), task (17.07%), eon card (7.32%), a notification card (7.32%), puffin card (6.1%), traffic card (3.66%) and sports card (1.22%). In the remainder of the paper, we will examine the dataset in detail.

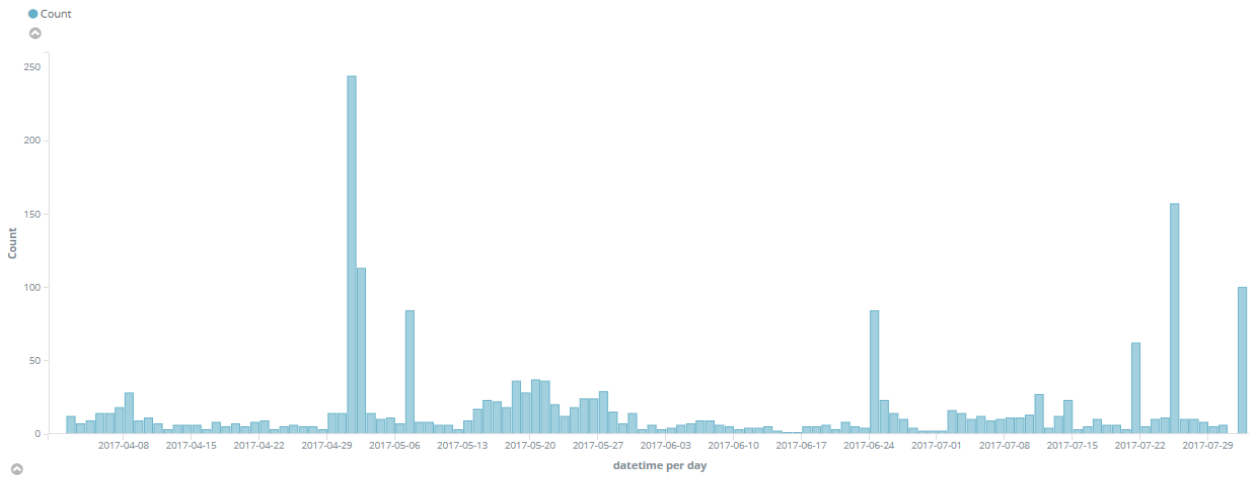

**Figure 2. Timeline of IVA usage history (April 1, 2017 ~ July 31, 2017)**

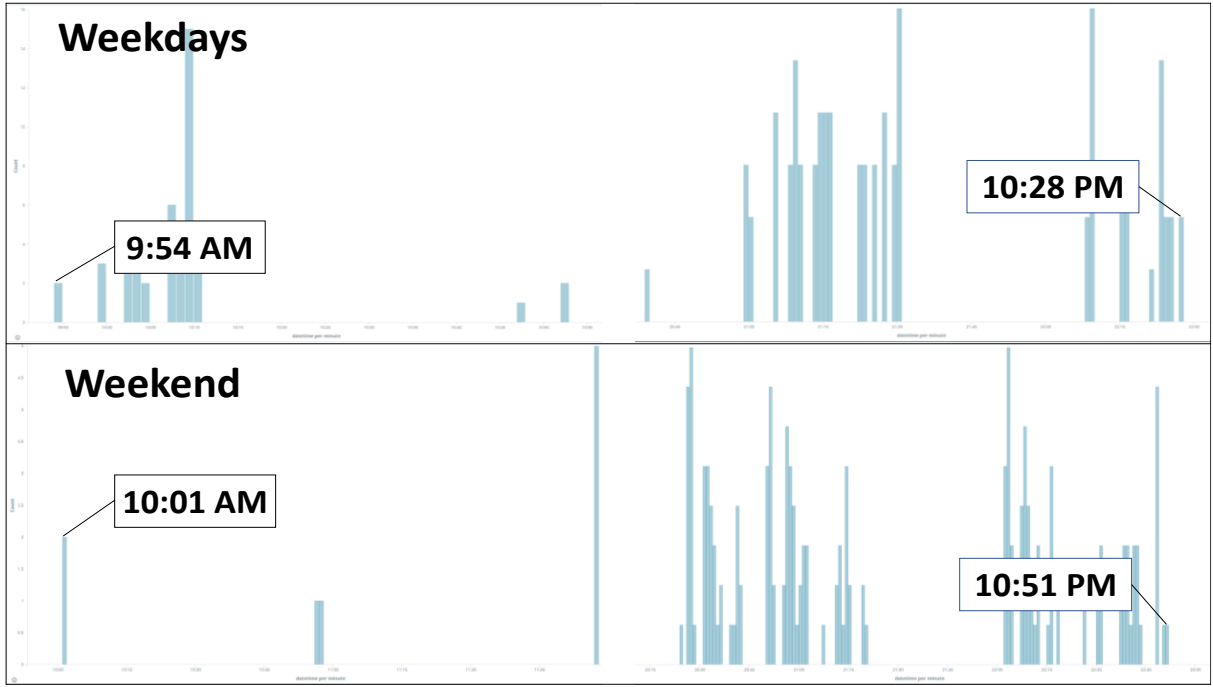

Figure 3. Two examples of daily IVA usage patterns

## 4 DETAILED FINDINGS

This section explains the results of data analysis using data visualization techniques. In this work, we utilized Elasticsearch for processing collected logs, and then visualized the data using a visualization toolkit, Kibana [9].

### 4.1 Timeline of IVA Usage History

**Figure 2** shows a timeline of usage history acquired after the participant interacted with Alexa cloud service through various Alexa-enabled devices. The timeline indicates that the participant utilized the IVA system frequently in daily life during a period, April 1 to July 31, 2017.

### 4.2 Daily IVA Usage patterns

We then considered detailed logs of each day. **Figure 3** shows two examples for showing daily IVA usage patterns. Based on the figure, we can identify the first and last usage time of Alexa-enabled devices.

During the experimental period, the participant usually started to communicate with Alexa between 8 AM to 10:10 AM. With an additional information such as alarm settings like shown in **Table 2**, it will increase a possiblility to understand user's lifestyle.

Table 2. Alarm history

| Date | Alarm setting |
|---|---|
| Apr 03, 2017 | Weekdays 7:50 AM |
| Apr 03, 2017 | Weekend 9:30 AM |
| Apr 05, 2017 | Everyday 5:30 PM |

In weekdays, between 6:30 AM and 8:00 AM, most people wake up, head into personal care such as showering and brushing teeth, and then head to work. Most people wake up later on the weekends. As people might expect, people head home to prepare and eat dinner around 5:00 PM [10].

As a result of data analysis, the participant usually did not use Alexa-enabled devices after 11 PM. Between 10:00 PM and midnight, people normally take a rest for the day and eventually going to sleep. By analyzing usage patterns with a reasonable inference, we may infer various life patterns such as wake up time, dinner time and sleep time.

### 4.3 User's Interests

Next, we focused on keywords detected from 'shopping item', 'puffin card' and 'sport card'. As a result, it revealed user's interests in daily life.

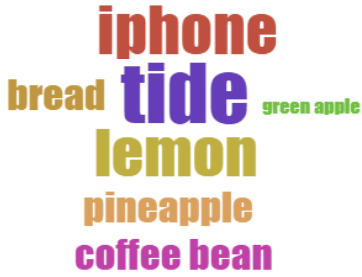

**Figure 4. Top 7 keywords of shopping list**

**Figure 4** shows the participant's shopping interests using a tag cloud. The tag cloud is an efficient visual representation of user's interests and preferences, because the importance (count) of each tag can be represented using variable font sizes. As shown in the figure, we found the fact that the participant added shopping items such as tide, iphone, lemon, pineapple, bread, coffee bean and green apple during the experimental period. Among detected keywords, the user especially interested in a few things including tide, lemon and iphone as expected shopping items.

**Figure 5. Personal interests on restaurants and sports**

**Figure 5** uncovers that the participant frequently tried to find Chiness and Japaness restaurants, and also shows a user's interest in sport.

### 4.4 User's Schedules

In addition, we tried to reconstruct the participant's to-do list and calendar using 'task' and 'eon card' log entries. For example, **Figure 6** shows a part of entire schedules acquired from cloud logs. The figure highlights several to-dos and schedules of May 2017, and we can identify that it includes business appointments and private affairs as well.

In the worst-case scenario, those kinds of private logs may be a critical point to stalk and harass a person, if a cyber attack can exfiltrate the data from service providers.

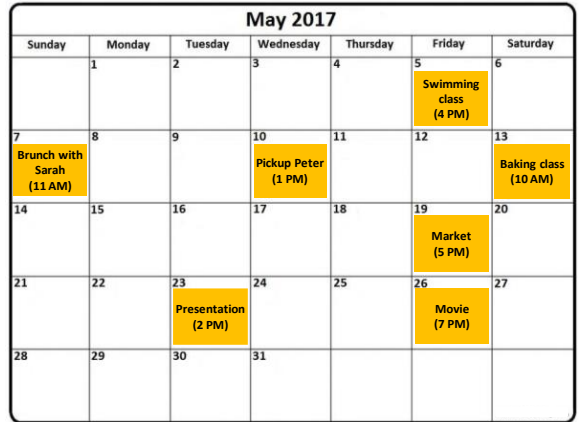

**Figure 6. Identified to-do and schedule list**

### 4.5 Driving Routes

In Alexa ecosystem, to check traffic condition, a user needs to configure starting and destination addresses for querying the information properly. Thereafter, it is possible to ask to IVA about traffic, for example "How is traffic?", "What is my commute?" and "What's traffic like right now?". As a result of those voice comands, location related data are stored in 'traffic card'. More specifically, 'Traffic card' normally includes the followings: location information (latitude and longitude) where the participant searched, a timestamp, source/destination addresses, distance, and time taken to reach a destination.

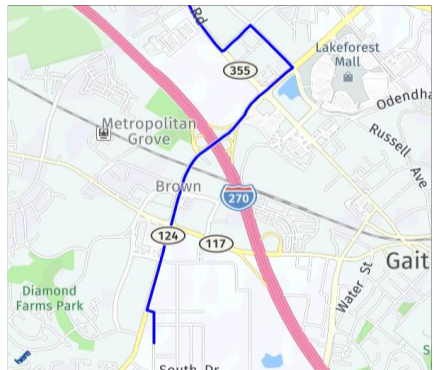

**Figure 7. A driving route between registered addresses**

**Figure 7** saved from Alexa cloud indicates a driving route between registered source and destination addresses. Addresses identified from IVA-related logs might include places closely associated with a user, such as home, school and workplaces.

## 5 DISCUSSIONS

### 5.1 User Privacy beyond Data Privacy

In Section 4, through data analysis using various visualization techniques, we found out forms and patterns stored within a participant's IVA-related data. As a result, it was possible to reveal a user's behaviors, usage patterns, interests and driving routes that the user searched. That is to say, we can infer and predict real world behaviors from cyber world data.

The modern IVA systems provide useful features like smartphones for users' daily life. For example, a user can utilize it conveniently with voice commands, for searching a keyword, playing music, adding to-do/shopping items, checking traffic conditions. In the context of data management, a user's smartphone data is basically stored into local storage devices, assuming that the user do not enable back-up features to cloud. In contrast, IVA services like Alexa tend to store personal information and usage history within cloud servers.

### 5.2 Suggestions for User Privacy

In this situation, it is necessary to consider about privacy issues and concerns. Based on our research findings that we explained in Section 4, we provide some suggestions for IVA users and vendors.

Firstly, IVA service users need to understand the risks of data breaches in this emerging environment. It is also necessary for IVA users to be aware of privacy risks. By creating strategies to protect user privacy, users should protect their personal information in cloud side. In details, users can periodically change passwords and delete usage logs from the cloud. For instance, Alexa users can either remove individual log entries, or delete entire usage history through de-registering devices.

Secondly, IVA vendor should try to protect customer's data for data and user privacy. For doing that, it is necessary for vendors to provide convenient methods that users can easily manage and delete their data.

## 6 CONCLUDING REMARKS

In recent days, cloud-based IoT devices are evolving rapidly and spreading widely in our lives. Many people are becoming accustomed to interacting with various IoT consumer products, such as intelligent virtual assistants. In these circumstances, lots of data are being produced in real time in response to user behaviors. Interestingly, a large amount of behavioral traces that include user's voice activity history with detailed descriptions can be stored in the remote servers.

Until now, there has been little research reported on analysis of intelligent virtual assistant related data collected from cloud servers. In this paper, we showed and categorized types of IVA-related data that can be collected from a popular IVA, Amazon Alexa. We then analyzed an experimental dataset from Amazon Alexa, and characterized several properties of a user's lifestyle and life patterns. Our results showed that it is possible to uncover new insights on personal information such as IVA usage patterns, user's interests and sleeping/wake-up patterns. The results presented in this work provide important implications for security and privacy threats to IVA vendors and users as well.

**Hyunji Chung** is a Ph.D. Candidate at the Graduate School of Information Security, Korea University, Seoul, South Korea. She is also a Guest Researcher at the National Institute of Standards and Technology's Computer Security Division, Gaithersburg, MD, USA since 2017. Her research interests include cloud-based IoT, security analysis and threat modeling.

**Sangjin Lee** received the Ph.D. degree from Korea University, Seoul, South Korea in 1994. He is a Professor of the School of Information Security, and also a Director of the Digital Forensics Research Center (DFRC), Korea University, Seoul, South Korea. He has served as a president, chair, committee member for many years in a variety of research societies, conferences and workshops. His research interests include security, digital forensics, incident response, steganography, cryptography and cryptanalysis.